\documentclass[preprint,aps,,amsmath,amssymb]{revtex4}

\usepackage{graphicx}
\usepackage{dcolumn}
\usepackage{bm}
\usepackage{color}

\begin{document}

\title{Parametric excitation of multiple resonant radiations from localized wavepackets}
\author{Matteo Conforti$^{1,*}$, Stefano Trillo$^{2}$, Arnaud Mussot$^{1}$, and Alexandre Kudlinski$^{1}$}

\affiliation{
$^1$PhLAM/IRCICA, CNRS-Universit\'e Lille 1, UMR 8523/USR 3380, F-59655 Villeneuve d'Ascq, France\\
$^2$Dipartimento di Ingegneria, Universit\`a di Ferrara, Via Saragat 1, 44122 Ferrara, Italy\\
email: matteo.conforti@univ-lille1.fr
}

\date{\today}

\maketitle

\begin{figure}[h]
\begin{center}
\includegraphics[width=12cm]{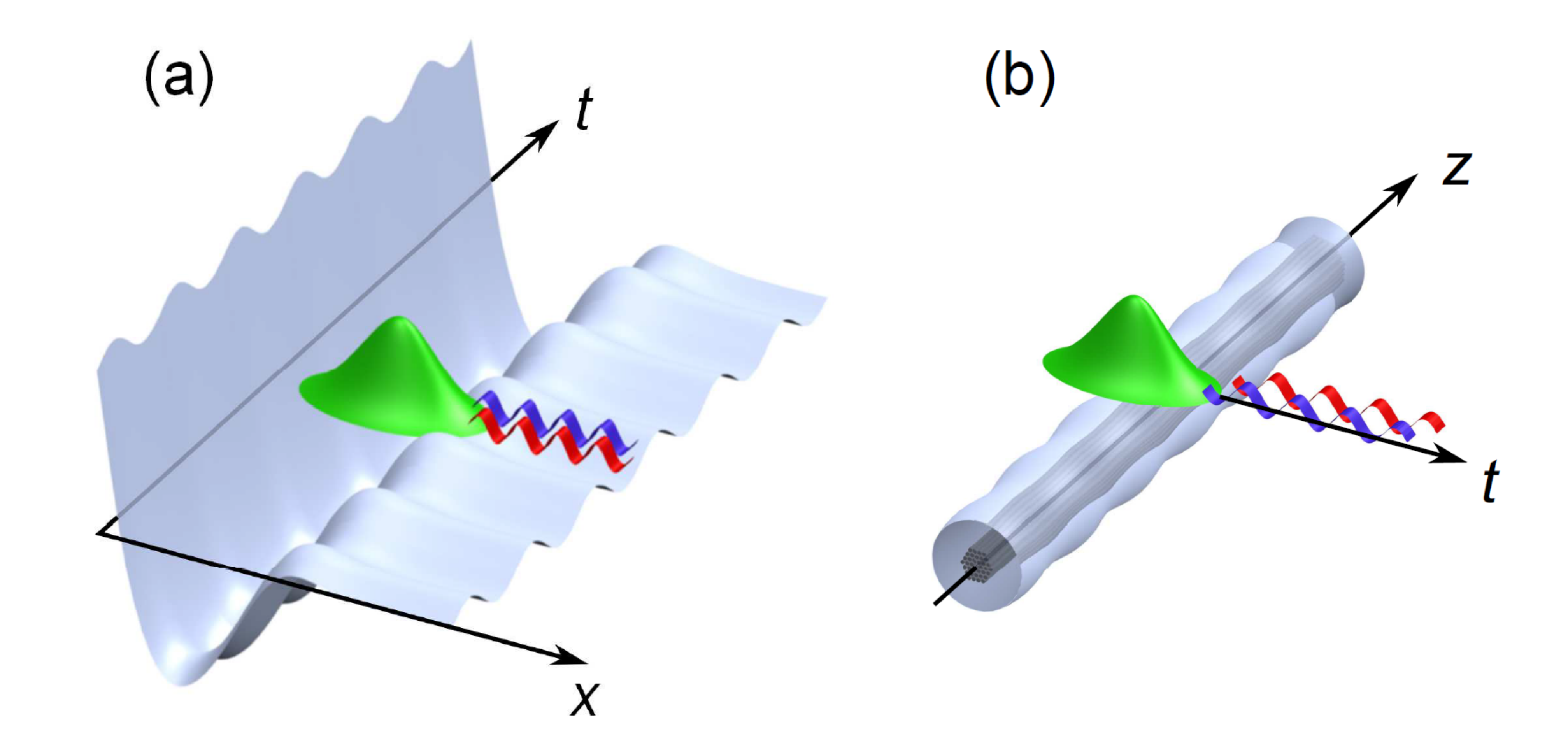}
    \end{center}
     \caption{{\bf Sketch of physically equivalent phenomena.} {\bf a}, coupling of bound states to continuum induced by generic potential which is oscillating in time. {\bf b}, fiber with oscillating diameter (dispersion) which induces wavepackets that are localized in time thanks to a self-induced potential given by the nonlinearity to couple into the radiation continuum at characteristic resonant frequencies.}
     \label{sketch}
\end{figure}

{\bf Fundamental physical phenomena such as laser-induced ionization \cite{ionization}, driven quantum tunneling \cite{tunneling}, Faraday waves \cite{Faraday1, Faraday4, Faraday2, Faraday3}, Bogoliubov quasiparticle excitations \cite{Casimir}, and the control of new states of matter \cite{Floquettopins1, Floquettopins2, Longhi13, BECmanage} rely on time-periodic driving of the system. A remarkable property of such driving is that it can induce the localized (bound) states to resonantly couple to the continuum (as illustrated in Fig. \ref{sketch}a). Therefore experiments that allow for enlightening and controlling the mechanisms underlying such coupling are of paramount importance.
{We implement such an experiment in a special optical fiber} characterized by a dispersion oscillating along the propagation coordinate, which mimics "time" (see Fig. \ref{sketch}b). The quasi-momentum associated with such periodic perturbation is responsible for the efficient coupling of energy from the localized wave-packets {(solitons in anomalous dispersion and shock fronts in normal dispersion)} sustained by the fiber nonlinearity, into free-running linear dispersive waves (continuum) at multiple resonant frequencies.
Remarkably, the observed resonances can be explained by means of a unified approach, regardless of the fact that the localized state is a soliton-like pulse or a shock front.}

Our experimental set-up realizes a platform for studying the coupling of localized states of a nearly conservative (Hamiltonian) system into radiation modes induced by periodic driving.
The localized states that we exploit are nonlinear (non-spreading) wavepackets of optical fibers, namely temporal solitons \cite{agrawal,DT09} and dispersive shock waves (DSWs) \cite{Rothenberg89,Wan07,Fatome14}. They share the property to have a well-defined wavenumber and group-velocity which are crucial for determining their resonances.
Their excitation is readily accessible by operating with opposite group-velocity dispersion (GVD, $\beta_2=d^2k/d\omega^2$): anomalous ($\beta_2<0$) for solitons in order to balance nonlinearity, whereas DSWs emerge from a gradient catastrophe occurring in the normal GVD  ($\beta_2>0$). Despite such difference, both are essentially described by a Hamiltonian model, namely the nonlinear Schr\"odinger equation (NLSE) \cite{agrawal,DT09}, where the propagation distance plays the role of evolution variable (usually time). This allows us to introduce periodic driving by employing photonic crystal fibers whose flexibility to engineer dispersion is fully exploited to have a longitudinal periodic GVD, thus realizing dispersion oscillating fibers (DOFs) \cite{droques13}.{We report clear evidence that this built-in and tailorable periodicity of the dispersion is responsible for the parametric excitation of radiation modes which are amplified out of quantum fluctuations at multiple resonant frequencies (even for a spatially harmonic variation). This thus constitutes a novel implementation of quasi-phase-matching \cite{QPM10}.} This is in marked contrast with radiation caused by standard perturbations such as third-order (or higher) dispersion \cite{ak95,skryabin10,Erkintalo12,Conforti13,Conforti14,Webb13}, which usually feature isolated resonances. Our results also clearly show that the parametric excitation of resonances is neither a prerogative of solitons nor of systems with periodic injection or extraction of energy such as fiber lasers \cite{kelly92} or lumped amplifier links \cite{smith92,turitsyn12} where dissipative resonances lead to the generation of so-called Kelly sidebands.


\section*{Results}
\textbf{Theory.} Perturbation theory allows us to predict the frequency detunings $\omega_{RR}$ of the resonant radiations (RR) that can be parametrically excited in a DOF. They turn out to be given by the roots of the following expression (see Supplementary information for a detailed derivation):
\begin{equation}\label{phase_matching}
\hat D(\omega_{RR})-k_{nl}=\frac{2\pi}{\Lambda}m,\;\;m=0,\pm 1,\pm2 ,\ldots
\end{equation}
{where $\hat D(\omega)=-\Delta k_1 \omega + \beta_2\omega^2/2 + \beta_3\omega^3/6 + \ldots$ is the average linear dispersion in the pump reference frame. $\beta_3 \equiv d^3k/d\omega^3$ and $\Delta k_1$ arises from the deviation of the actual group-velocity from the natural one \cite{Conforti13}). In Eq.~(\ref{phase_matching}), $k_{nl}$ is a well defined nonlinear wavenumber fixed by power. For a bright soliton $k_ {nl}=\gamma P/2$, where $P$ and $\gamma$ are the soliton peak power and the fiber nonlinearity, whereas for a shock wave $k_{nl}=-\gamma P_b$, being $P_b$ the power of the background field over which the RR propagates \cite{Conforti14}}. Equation (\ref{phase_matching}) generalizes previously proposed formulas \cite{abdullaev94,pelinovsky04} and expresses momentum conservation: it states that the difference between the momentum of the linear waves and the momentum generated by the nonlinear pump must be equal to the virtual momentum carried by the periodic modulation of the dispersion.

\begin{figure}[h]
\begin{center}
\includegraphics[width=\textwidth]{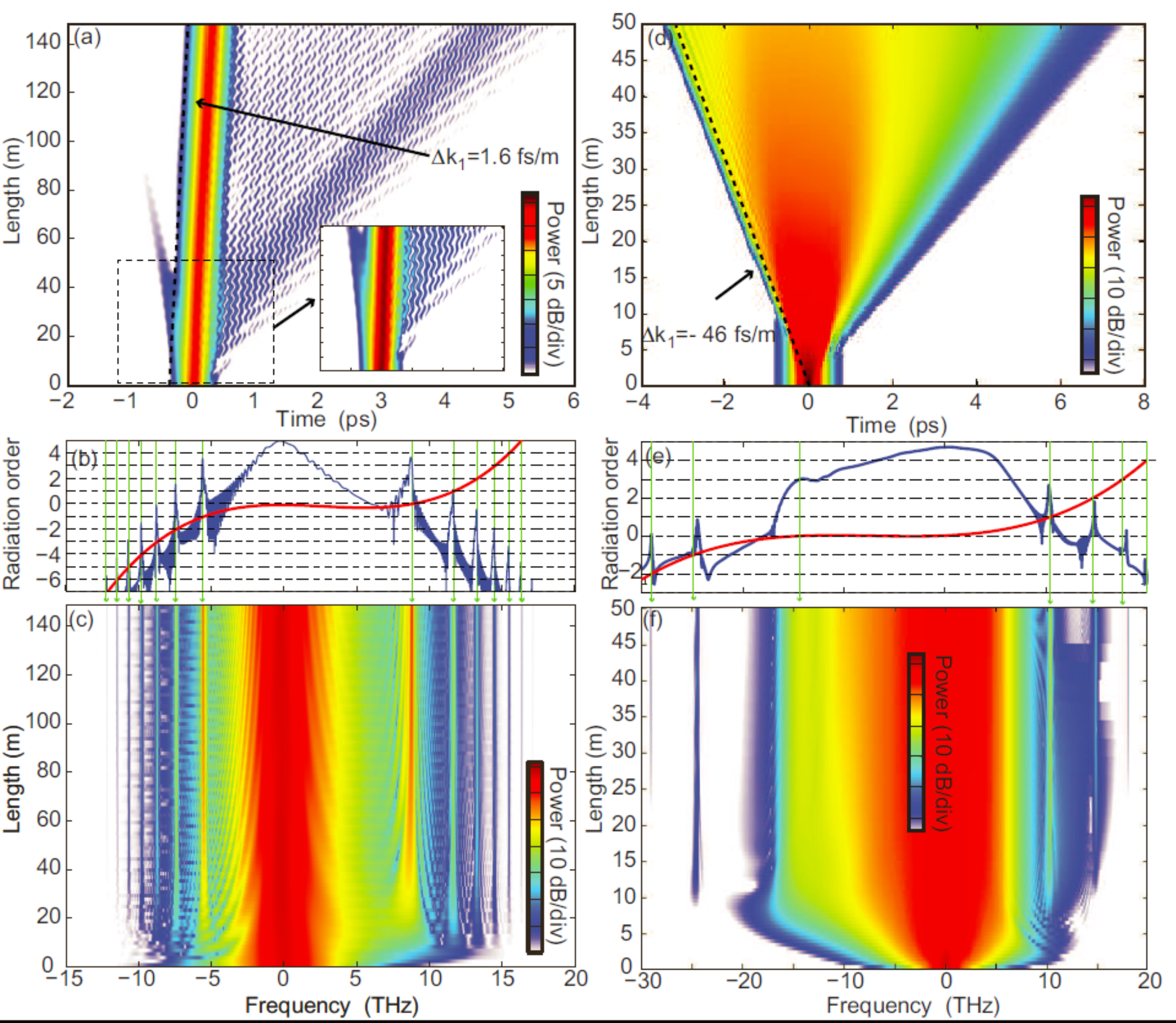}
    \end{center}
     \caption{{\bf Parametrically excited multiple resonances at fixed power (numerical modeling).} Left panels a,b,c refer to the soliton pumped configuration, whereas right panels d,e,f refer to the shock configuration.
{\bf a,d}: colormap of the temporal evolution of power (log scale) along the fiber; {\bf b,e}: output spectra superimposed on the dispersion relation $\hat D(\omega)-k_{nl}$ (red curve). The vertical green lines correspond to the resonances determined by the graphical solution of Eq. (\ref{phase_matching}), i.e. the crossing between the dispersion curve and the dashed horizontal lines that corresponds to integer multiples of quasi-momentum $2\pi/\Lambda$; {\bf c,f}: colormap evolution of the spectrum along the fiber. All the results are obtained by numerical simulation of the NLSE (see methods), with periodic GVD and $\beta_3$ as additional parameter. }
\label{example}
\end{figure}

We verified that Eq. (\ref{phase_matching}) describes accurately the parametric excitation of RR in both GVD regimes. {A clear illustration} of this phenomenon is provided in Fig. \ref{example}, which shows the simulated evolution of a hyperbolic secant pulse ($150$ fs duration, peak power $P$) based on the NLSE with included periodic GVD and dispersion slope $\beta_3$ (see methods).
Figure \ref{example}a displays the time domain evolution of a pulse with $P=15$ W, corresponding to a nearly fundamental soliton launched in a 150 m long DOF of period $\Lambda=5$ m. Strong radiation traveling slower than the soliton becomes visible beyond the activation distance $z_a \sim 20$ m, corresponding to maximum pulse compression. In the spectral domain, the radiation modes correspond to several distinct and well defined frequencies [see Fig. \ref{example}b,c] that agree with the prediction based on Eq. (\ref{phase_matching})(vertical green lines). The peak detuned by $\sim 9$ THz from the pump turns out to be the standard RR [$m=0$ in Eq. (\ref{phase_matching})], whereas other twelve peaks originate from the periodic perturbation [$m\neq0$ in Eq. (\ref{phase_matching})].

When pumping in the normal GVD regime, we selected a higher peak power and duration ($P=100~W$, $280~fs$) in order to access wave-breaking, and a shorter modulation period ($\Lambda=0.5~m$) in order to have resonances in a realistic frequency span. In this case, the localized state is the shock front that emerges over the pulse leading edge after the breaking and activation distance $z \sim 8$ m, which is clearly visible in Fig. \ref{example}d. Also in this regime, we identify a first spectral peak at $-15$ THz in the anomalous GVD region [see Fig. \ref{example}e,f] as the standard RR ($m=0$ mode) due to $\beta_3$, while other five peaks are clearly visible. These are parametrically excited resonant modes that correspond to $m=\pm 1, \pm 2$, $3$ which position are also well predicted by Eq. (\ref{phase_matching})(see vertical green lines).

We emphasize that in both configurations, the $m=0$ isolated resonance lies in the opposite GVD regime with respect to the localized wavepackets that generate it, and disappear for $\beta_3=0$. {Conversely, the parametric resonances with $m \neq 0$ lie in both normal and anomalous GVD regimes. They also survive the absence of higher-order dispersion ($\beta_n=0$, $n \ge 3$), and their number is infinite for a harmonic perturbation.} In practice, their actual number is limited because their amplitude decreases with the frequency shift and they become weaker than the pulse spectrum envelope and/or the noise background.
{Note also that it is important to take into account the group-velocity deviation $\Delta k_1$ [see Fig. \ref{example}a,d] in the $\hat D$ term of Eq.~(\ref{phase_matching}) in both configurations, in order to accurately predict the actual values of the resonances.}

\begin{figure}[h]
\begin{center}
\includegraphics[width=\textwidth]{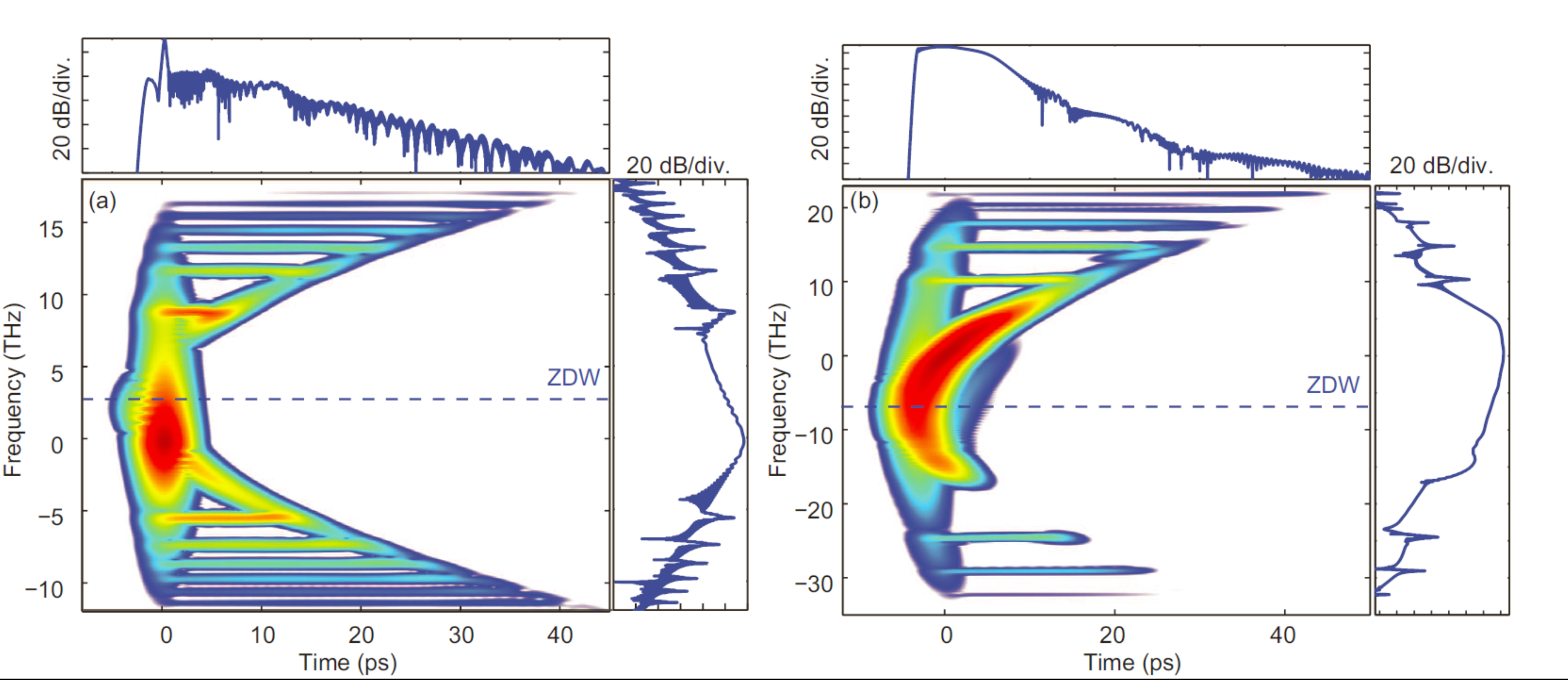}
    \end{center}
     \caption{{\bf Spectrograms corresponding to evolutions in Fig. \ref{example}  (numerical modeling).} a: soliton configuration. b: shock wave configuration. The spectrograms represent the spectra over gated time intervals and are computed at the output of each fibers (for the evolution along the fibers, see the additional multimedia files). ZDW: zero-dispersion wavelength.}
     \label{spectro}
\end{figure}

The nature of coupling process is further clarified by the spectrograms corresponding to Fig. \ref{example}, which are displayed in Fig. \ref{spectro}. In both cases, the pump wave-packets remain clearly localized in both temporal and spectral domains, while shedding energy into radiation modes, which disperse in time while remaining at the frequencies predicted by the resonances in Eq. (\ref{phase_matching}). We have found that the localized pumps experience temporal and spectral breathing with the period of the perturbation and refill the radiation at each cycle of spectral broadening (see multimedia files). In spite of this transfer of energy the radiation damping of the pump is small, thus confirming the metastable nature of these wave-packets \cite{Weinstein01}.

\begin{figure}[h]
\begin{center}
\includegraphics[width=\textwidth]{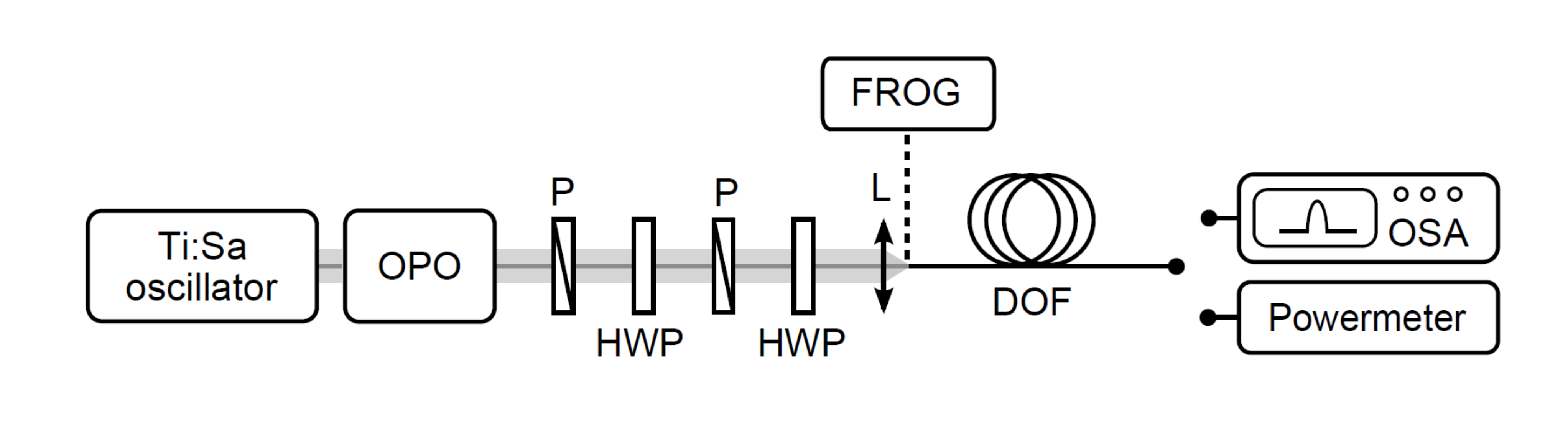}
    \end{center}
     \caption{{\bf Experimental set-up} P: polarizer; HWP: half-wave plate.} \label{fig4}
\end{figure}

\begin{figure}[h]
\begin{center}
\includegraphics[width=\textwidth]{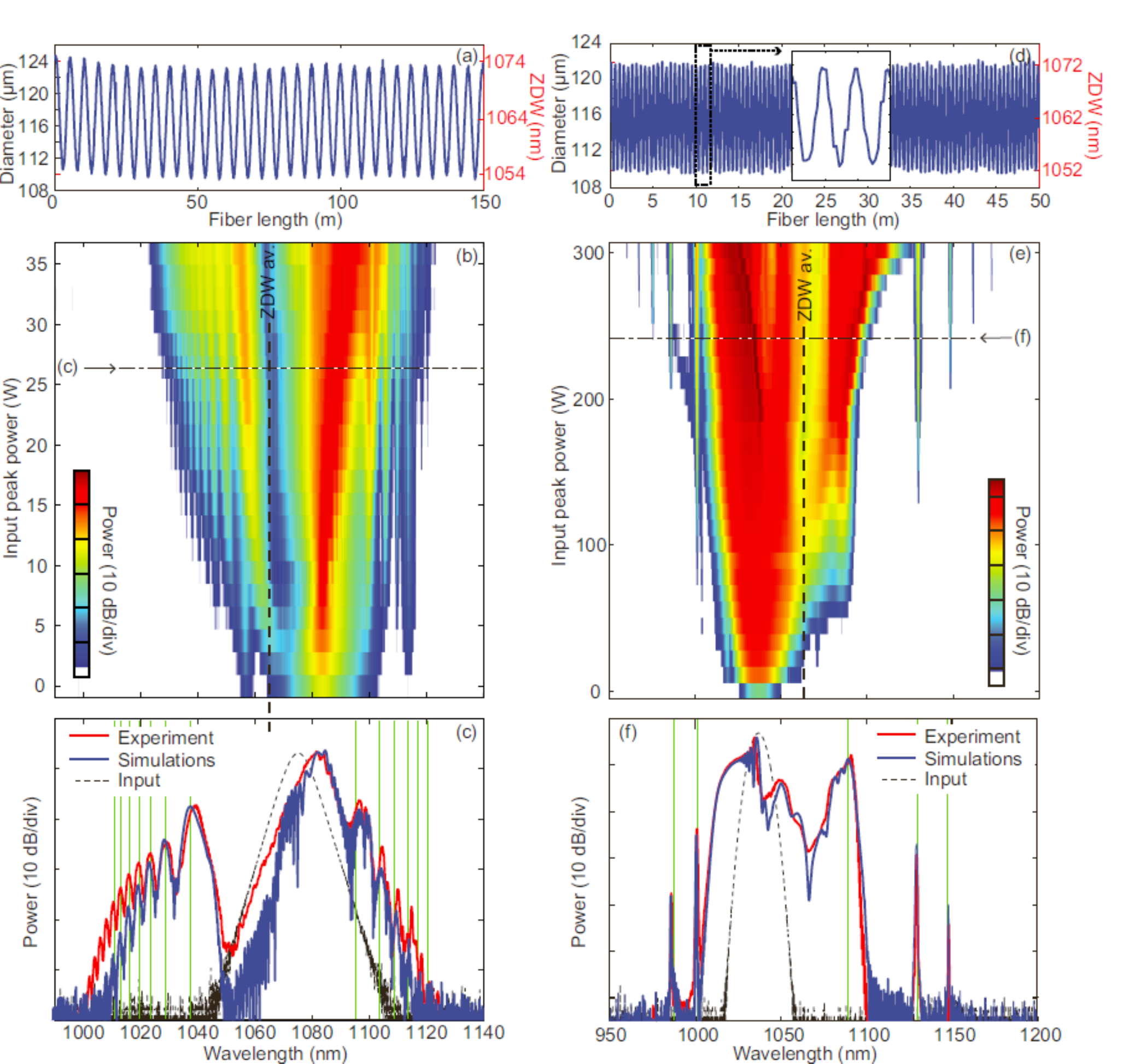}
    \end{center}
     \caption{{\bf Parametrically excited multiple resonances from a fundamental soliton (left panels) and from a DSW (right panels): experimental results.} {\bf a,d}: Longitudinal evolution of the DOF diameter (left vertical axis) and consequent zero-dispersion wavelength (ZDW, right vertical axis) used for the soliton (a) and dispersive shock waves (d) experiments. {\bf b,e}: Experimental power maps showing the development of asymmetric sidebands with increasing pump peak power. {\bf c,f}: Comparison between experimental (red lines) and simulated spectra (blue lines) for a pump peak power of 26.4 W in the soliton case (c) and of 234 W in the dispersive shock wave case (f). Vertical green lines depict the resonance from Eq. (\ref{phase_matching}). }
     \label{experiment}
\end{figure}

\textbf{Experiments.} We have designed two experiments in order to observe the parametrically excited resonances from both solitons and DSWs. The experiment, sketched in Fig. \ref{fig4} and detailed in methods, simply consists of a femtosecond laser pulse launched in DOFs, whose parameters are close to those used in the illustration examples in Figs. \ref{example}-\ref{spectro}. Due to experimental constraints, only the initial pulses were slightly different. In a first experiment, a 150 fs pulse centered at 1075 nm was launched in a 150 m-long DOF with a modulation period of 5 m. Figure \ref{experiment}a shows the longitudinal profile of the fiber diameter measured during the draw process (left axis) and the corresponding calculated zero dispersion wavelength (ZDW, right axis). At the pump wavelength of 1075 nm, the average dispersion is slightly anomalous, so that the pump pulse is able to excite a near-fundamental soliton for peak powers in the order of a few tens of watts. An experimental power map, representing the output spectrum recorded for increasing pump peak power, is plotted in Fig. \ref{experiment}b. It shows that, for increasing pump peak power, the spectrum rapidly evolves from a hyperbolic secant shape to a much more structured and highly asymmetric one containing more and more sharp  spectral resonances. More precisely, these discrete spectral sidebands corresponds to the parametric excitation of the RR that stems from the periodic variation of the second order dispersion. {Numerical simulations of the generalized NLSE (see methods and supplementary information), without any free parameter, also reproduce these features with an excellent quantitative agreement. This can be seen in Fig. \ref{experiment}c, where the experimental spectrum obtained for a pump peak power of 25 W (red line) is compared with the simulated one (blue line).} In addition, it is worth noting that the theory is quite robust since the observed peaks are accurately predicted by solutions of Eq. (\ref{phase_matching}) (green vertical lines) in these experiments.

The second experiment presented here is devoted to investigating radiating DSWs. The DOF is shorter (50 m) as well as the modulation period (0.5 m, see longitudinal profile in Fig. \ref{experiment}d). It has been pumped  with 280 fs pulses. The pump wavelength was tuned to 1037 nm so that it lies in the normal average dispersion region required to generate a shock wave from a few hundreds of watts of peak power (see additional experimental details in supplementary information). Figure \ref{experiment}e shows the experimental power map. Starting at about 50 W, a RR peak is generated across the average zero dispersion wavelength. For increasing pump peak powers, additional discrete spectral sidebands corresponding to parametrically excited RR appear, similarly to the soliton case. The DOF period $\Lambda$ being 10 times shorter in this case, the spacing between two adjacent peaks is $\sqrt{10}$ times larger than in the soliton case, as expected from Eq. (\ref{phase_matching}). These results are again in excellent agreement with numerical simulations using a generalized NLSE and with Eq. (\ref{phase_matching}), as shown in Fig. \ref{experiment}f and in supplementary information.

\section*{Conclusions}
{To summarize, we have demonstrated that localized states (either a soliton or a dispersive shock wave) can efficiently transfer energy to multiple resonant radiations at different frequencies, as a result of the quasi-momentum associated to a DOF. Our experimental results, supported by numerical simulations and by the perturbative analysis that leads to Eq. (\ref{phase_matching}), prove that a DOF is a very simple and highly tailorable platform allowing to study the periodicity-induced coupling of nonlinear bound states to the radiation continuum, which is a general feature in systems driven by a time-periodic Hamiltonian. The DOF platform can also be successfully used to study how this coupling process due to higher order dispersion develop in the presence of a train of random solitons arising from spontanous modulation instability or in driven-damped deformations of Hamiltonian systems such as those describing passive ring resonators. 
 }

\section*{Methods}

\textbf{Simulation:} The results illustrated in Figs. \ref{example}-\ref{spectro} have been obtained from numerical integration of the following NLSE for the electric field envelope $E=E(z,t)$ propagating along the fiber \cite{agrawal}
\begin{equation}
\label{nls}
\nonumber i\frac{\partial E}{\partial z} -\frac{\beta_2(z)}{2}\frac{\partial^2 E}{\partial t^2}+\frac{\beta_3}{6}\frac{\partial^3 E}{\partial t^3} + \gamma |E|^2 E = 0,
\end{equation}
{Equation~\ref{nls} is well known to maintain conservative (Hamiltonian) structure despite the periodic perturbation embedded in the term  $\beta_2(z)=\beta_2+\delta\beta_2\sin(2\pi z/\Lambda)$ (where $\delta \beta_2$ is the perturbation amplitude around the average GVD $\beta_2$), and the third-order dispersion $\beta_3$ ( which plays a significant role in experiments performed by pumping close to the ZDW, as in our case).} We have employed the following values of the parameters that arise from fiber characterization: $\gamma=10$ (W km)$^{-1}$, $\delta \beta_2=1.2$ ps$^2$/km, $\beta_2=-1.2$ ps$^2$/km and $\beta_3=0.0716$ ps$^3$/km for the soliton configuration, whereas  $\beta_2=2.9$ ps$^2$/km and $\beta_3=0.0645$ ps$^3$/km for the DSW configuration. The nonlinear term in Eq. (\ref{nls}) acts as a self-induced potential that allows for the existence of localized states in the form of bright solitons for $\beta_2<0$ (anomalous dispersion) or dispersive shock waves for $\beta_2>0$ (normal GVD), whose leading front is, in the unperturbed case, locally a dark soliton \cite{Conforti14}.
{The radiating soliton excited in Fig. \ref{example}a,b,c is characterized by a soliton number $N = \sqrt{T_0^2 \gamma P/|\beta_2|}$ slightly higher than $1$,
so that the pulse undergoes temporal compression and hence spectral broadening. It thus acts as an effective seed for the parametric excitation of RR.}
We emphasize that the periodic perturbation is nearly resonant with the soliton period ($\simeq 10.5~m$), which distinguishes our regime from non-radiating guiding-center or average solitons \cite{hk91,Pelinovsky03}, which are normally seen when the periodic perturbation is much faster than the soliton period. {In the shock case, the equivalent quantity $N = \sqrt{T_0^2 \gamma P/\beta_2}$ is much larger than unity ($N \gg 1$) so that the nonlinearity drives the pulse towards the gradient catastrophe that causes the formation of the shock front \cite{Fatome14}.}

Numerical simulations of Eq. (\ref{nls}) have been performed by using the split step Fourier method with a temporal resolution of 5 fs, $2^{14}$ points and a spatial resolution ranging from 0.025 m to 0.1 m for the shock and the soliton configurations, respectively. 

When directly comparing the numerics with the experimental data (results in Fig. \ref{experiment}), we have accounted also for secondary effects in the fiber such as losses, Raman scattering, self-steepening, and fourth-order dispersion by making use of a generalized NLSE (Eq. (2.3.36) in Ref. \cite{agrawal}, also reported explicitly in the Supplemental information). However, we have verified that the impact of such effects on the pulse evolution over the length and temporal scales involved in the experiment is really minor, and most of all does not affect the resonances.

\textbf{Experiments:} Experiments have been performed exploiting a Ti:Sa oscillator delivering 140 fs full width at half maximum (FWHM) near transform limited pulses. They are sent into an optical parametric oscillator (OPO) allowing to generate tunable slightly chirped femtosecond pulses. The output beam then passes through a combination of two polarizers and two half-wave plates in order to carefully adjust the polarization state and pump power simultaneously. It is launched in the DOF with an aspherical lens. The pulses were characterized with a {frequency resolved optical gating (FROG)} system, before being injected into the DOF.
For the soliton experiment, the OPO was bypassed so that pulses directly coming  from the Ti:Sa oscillator were  used. They were centred at 1075 nm, and they were measured at 150 fs FWHM at the DOF input (i.e. after the launch lens) with a small chirp $C=0.35$ (as defined in \cite{agrawal}). For the dispersive shock wave experiment, the OPO was used and tuned to 1037 nm. The pulses at the DOF input were measured at 280 fs FWHM, with a chirp $C=1.24$.
Spectra out of the DOF were acquired with an optical spectrum analyzer with a resolution of 0.2 nm. The output power was measured with a power-meter and the input power was deduced knowing the DOF total attenuation. It was cross-checked by cutting the DOF at the end of the experiment and measuring the power out of the 0.5 m DOF initial section.

\section*{Acknowledgements}
This work was partly supported by the Agence Nationale de la Recherche through the ANR TOPWAVE project, the Labex CEMPI and Equipex FLUX through the "Programme d'Investissement d'Avenir", by the French Ministry of Higher Education and Research, the Nord-Pas de Calais Regional Council and Fonds Europ\'{e}en de D\'{e}veloppement R\'{e}gional (FEDER) through the "Contrat de Projets Etat R\'{e}gion (CPER) 2007-2013" and the "Campus Intelligence Ambiante (CIA)", and by the by the Italian Ministry of University and Research (MIUR) under Grant PRIN 2012BFNWZ2. We are grateful to Francois Copie for producing Figure 1.

\section*{Authors contributions}
A.K. carried out experiments. The development of analytical
tools was carried out by M.C. and S.T. and simulations were performed by M.C. and A.M.
All authors conceived the idea of this work, participated in the analysis and interpretation of the results and in the
writing of the paper.

\section*{Additional information}
{\bf Supplementary information} accompanies this paper at http://www.nature.com/
scientificreports
\\
{\bf Competing financial interests:} The authors declare no competing financial interests.

\clearpage

\clearpage 

\section{ Supplementary information}

\subsection{Derivation of the resonances}
In order to derive the equation of the resonances [Eq. (1) of the paper], we start from the following conservative nonlinear Schr\"odinger equation (NLSE)
\begin{align}
&i\partial_z u + d(i\partial_t)u -\frac{\delta \beta_2(z)}{2}\partial^2_t u + \gamma|u|^2 u=0,\label{gNLS}\\
&d(i\partial_t)=\sum_{n\geq 2}\frac{\beta_n}{n!}(i\partial_t)^n,\;\;\; D(\omega)=\sum_{n \geq 2}\frac{\beta_n}{n!}\omega^n.
\end{align}
where $t$ stands for time measured in the frame moving at natural group velocity of light $V_g=dk/d\omega^{-1}$ \cite{agrawal}, $\gamma$ is the nonlinear coefficient, $\beta_n=d^nk/d\omega^n$ is the $n$-th order average dispersion, $d(i\partial_t)$ and $D(\omega)$ are the average dispersion operators in time and frequency domain (Fourier transform), respectively. In Eq. (\ref{gNLS}) the average group-velocity dispersion (GVD) $\beta_2$ is perturbed by the periodic modulation $\delta \beta_2 (z)$ and by higher-order dispersive effects ($\beta_n$, $n \ge 3$).

We consider a localized wave-packet (either a soliton or a shock front) that moves with characteristic velocity $V_s=1/\Delta k_{1}$, or in other words a solution $u_s(z,t)=A(\tau) \exp(iqz)$ of the unperturbed Eq. (\ref{gNLS}) which has nonlinear wavenumber $q$ and is stationary in the reference frame $\tau=t-\Delta k_1 z$. In terms of this new variable Eq. (\ref{gNLS}) reads as:
\begin{align}
&i\partial_z u + \hat d(i\partial_\tau)u -\frac{\delta \beta_2(z)}{2}\partial^2_\tau u + \gamma|u|^2u=0,\label{NLS}\\
&\hat d(i\partial_\tau)=\sum_{n\geq 2}\frac{\beta_n}{n!}(i\partial_\tau)^n-\Delta k_1 i\partial_\tau, \;\;\;  \hat D(\omega)=\sum_{n \geq 2}\frac{\beta_n}{n!}\omega^n-\Delta k_1\omega.
\end{align}

%
When the perturbation is effective, the solution can be decomposed into the localized wave and radiation modes $g(z,\tau)$
\begin{equation}
u(z,\tau)=A(\tau)e^{iqz}+g(z,\tau).
\end{equation}

Assuming the radiation modes to be weak ($|g| \ll |A|$), the linearization of Eq. (\ref{gNLS}) around the localized wave yields
\begin{equation}\label{lin}
i\partial_z g+ \hat d(i\partial_\tau)g-\frac{\delta \beta_2 (z)}{2}\partial_{\tau}^2g+ \gamma \left(u_s^2g^*+2|u_s|^2g\right)=\frac{\delta\beta_2(z)}{2}\partial_t^2 u_s-\hat d_H(i\partial_\tau)u_s,
\end{equation}
where $\hat d_H(i\partial_\tau)=\sum_{n\geq 3}\frac{\beta_n}{n!}(i\partial_\tau)^n$
accounts for the perturbation due to higher-order dispersive terms. The RHS of Eq. (\ref{lin}) represents the driving force for the evolution of the radiation modes. Importantly, the effective wavenumber of this forcing is not simply the wavenumber of the localized wave-packet but is also affected by the quasi-momentum associated with the periodic modulation of the second order dispersion.

Let us focus first on the free evolution of the system, described by the LHS of Eq. (\ref{lin}). Without loss of generality, the radiation can be searched in the form
\begin{equation}\label{g}
g(z,t)=\left[a(z)e^{i(kz-\omega \tau)}+b^*(z)e^{-i(kz-\omega \tau)}\right]\exp\left[i q z+i\frac{\omega^2}{2}\int_0^z\delta\beta_2(s)ds\right].
\end{equation}
We find the following system that rules the free evolution [i.e. neglecting the forcing term corresponding to RHS of Eq. (\ref{lin})] of the dispersive waves
\begin{equation}
i\partial_z\left[
\begin{array}{c}
a \\b
\end{array} \right]
+
\left[
\begin{array}{ c c}
\hat D(\omega)-k-q+2\gamma |A|^2 & \gamma A^2\exp\left[-i\omega^2\int_0^z\delta\beta_2(s)ds\right]\\
-\gamma (A^*)^2\exp\left[i\omega^2\int_0^z\delta\beta_2(s)ds\right] & -\left( \hat D(-\omega)+k-q+2\gamma |A|^2\right)
\end{array} \right]\left[
\begin{array}{c}
a \\b
\end{array} \right]=0.
\end{equation}
The dispersion relation of the linear waves $k=k(\omega)$ can be found by setting the determinant of the matrix equal to zero. In terms of the odd and even dispersive contributions ($\hat D_o(\omega)=\hat D(\omega)-\hat D(-\omega)$ and $\hat D_e(\omega)=\hat D(\omega)+\hat D(-\omega)$, respectively), we find
\begin{equation}\label{disp}
k_{\pm}(\omega)=\frac{\hat D_o(\omega)}{2}\pm\frac{1}{2}\sqrt{\left[\hat D_e(\omega)-2q +2\gamma |A|^2 \right]\left[\hat D_e(\omega)-2q +6\gamma |A|^2 \right]}.
\end{equation}

For the soliton, Eq. (\ref{disp}) holds true with $A=0$ since radiation is not temporally overlapped with the soliton (it grows on soliton tails where $A$ exponentially vanishes) and $q=\gamma P/2$ \cite{agrawal}, being $P$ the peak power of the soliton. Therefore we obtain
\begin{equation}
k(\omega)=\hat D(\omega)-\gamma\frac{P}{2}.
\end{equation}

For the shock wave, the radiation modes are amplified out of noise close to the leading edge of the dispersive shock wave. Therefore they propagate over the flat-top background $A_0$ (power $P_b=|A_0|^2$), which develops as a result of the steepening of the pulse edges \cite{Conforti14}, which is in turn responsible for the shock formation. In this case we can set $q=\gamma P_b$, i.e. the wavenumber of the gray soliton associated to the leading edge of the dispersive shock wave.
In this case Eq. (\ref{disp}) yields
\begin{equation} \label{dispshock}
k_\pm(\omega)=\frac{\hat D_o(\omega)}{2}\pm\frac{1}{2}\sqrt{\hat D_e(\omega)\left[\hat D_e(\omega)+4\gamma|A_0|^2 \right]}.
\end{equation}
In the shock case the presence of the background generates two symmetric branches of the dispersion relation. This fact accounts for the four wave mixing between the dispersive waves and the pump. Usually the amplitude of the two symmetric waves ($a$ and $b$) are orders of magnitude different, so that only one branch of dispersion relation ($k_+$ in our case) turns out to be relevant. Under the hypothesis $|A_0|^2 \ll |\hat D_e|$, we can expand in Eq. (\ref{dispshock}) the square root  to obtain, for the relevant branch,
\begin{equation}
k(\omega)=\hat D(\omega)+\gamma P_b.
\end{equation}

The forcing term $F$ arising from the RHS of Eq. (\ref{lin}) which is effective for the growth of the radiation modes with complex amplitudes $a$ and $b$, turns out to be
\begin{equation} \label{forcing}
F=\frac{\delta\beta_2(z)}{2}\exp\left[-i\frac{\omega^2}{2}\int_0^z\delta\beta_2(s)ds\right]\partial_t^2A-\exp\left[-i\frac{\omega^2}{2}\int_0^z\delta\beta_2(s)ds\right]\hat d_H(i\partial_\tau)A.
\end{equation}

By considering a periodic $\delta \beta_2(z)$ with period $\Lambda$, we can expand the perturbation in Fourier series of the form $\delta\beta_2(z)=\sum_l c_l e^{i l\frac{2\pi}{\Lambda} z}$,
and consequently expand the exponential in Eq. (\ref{forcing}) as
\begin{align}
&\exp\left[-i\frac{\omega^2}{2}\int_0^z\delta\beta_2(s)ds\right]=\sum_m d_m e^{im\frac{2\pi}{\Lambda} z}. \label{exp}
\end{align}
Therefore we cast Eq. (\ref{forcing}) in the form
\begin{equation}
F=\partial_\tau^2A\sum_{l,m} c_n d_m e^{i(l+m)\frac{2\pi}{\Lambda} z}-\hat d_H(i\partial_\tau)A\sum_m d_n e^{im\frac{2\pi}{\Lambda} z},
\end{equation}
which allow to recognize two different driving terms for the growth of radiation modes. The first one comes from the modulation of the GVD, whereas the second one from higher-order dispersive terms. Importantly,  even in the case of sinusoidal modulation, the exponential term (\ref{exp}) generates an infinite set of Fourier harmonics.

Coupling of energy into the radiation modes (dispersive waves) efficiently occurs when their wavenumber $k(\omega)$ equals the wavenumber of the forcing term. This leads to the following resonance condition (quasi-phase-matching)
\begin{equation}\label{phase}
k(\omega)=\frac{2\pi}{\Lambda}m,\;\;m=0,\pm 1,\pm 2,\ldots
\end{equation}
or, equivalently, in the form of Eq. (1) of the paper:
\begin{equation}\label{phase_matching}
\hat D(\omega)-k_{nl}=\frac{2\pi}{\Lambda}m,\;\;m=0,\pm 1,\pm2 ,\ldots
\end{equation}
where $k_{nl}=\gamma P/2$ and $k_{nl}=-\gamma P_b$, for the soliton and the shock configuration, respectively.

Equation (\ref{phase}) retains its validity for any general dispersion profile. However, in our experiments the dispersive operator can be truncated to the first correction to GVD, i.e. third-order dispersion $\beta_3$, whereas all the higher-order dispersive terms can be safely neglected. In this case, Eq. (\ref{phase}) can be cast, for the soliton configuration, in the final form
\begin{equation}
\frac{\beta_3}{6}\omega^3+\frac{\beta_2}{2}\omega^2-\Delta k_1\omega-\gamma\frac{P}{2}=\frac{2\pi}{\Lambda}m,
\end{equation}
whereas for the shock configuration, we obtain
\begin{equation}
\frac{\beta_3}{6}\omega^3+\frac{\beta_2}{2}\omega^2-\Delta k_1\omega+\gamma P_b=\frac{2\pi}{\Lambda}m.
\end{equation}

\subsection{Properties of the dispersion oscillating fibers}
\begin{figure}[h]
\begin{center}
\includegraphics[width=\textwidth]{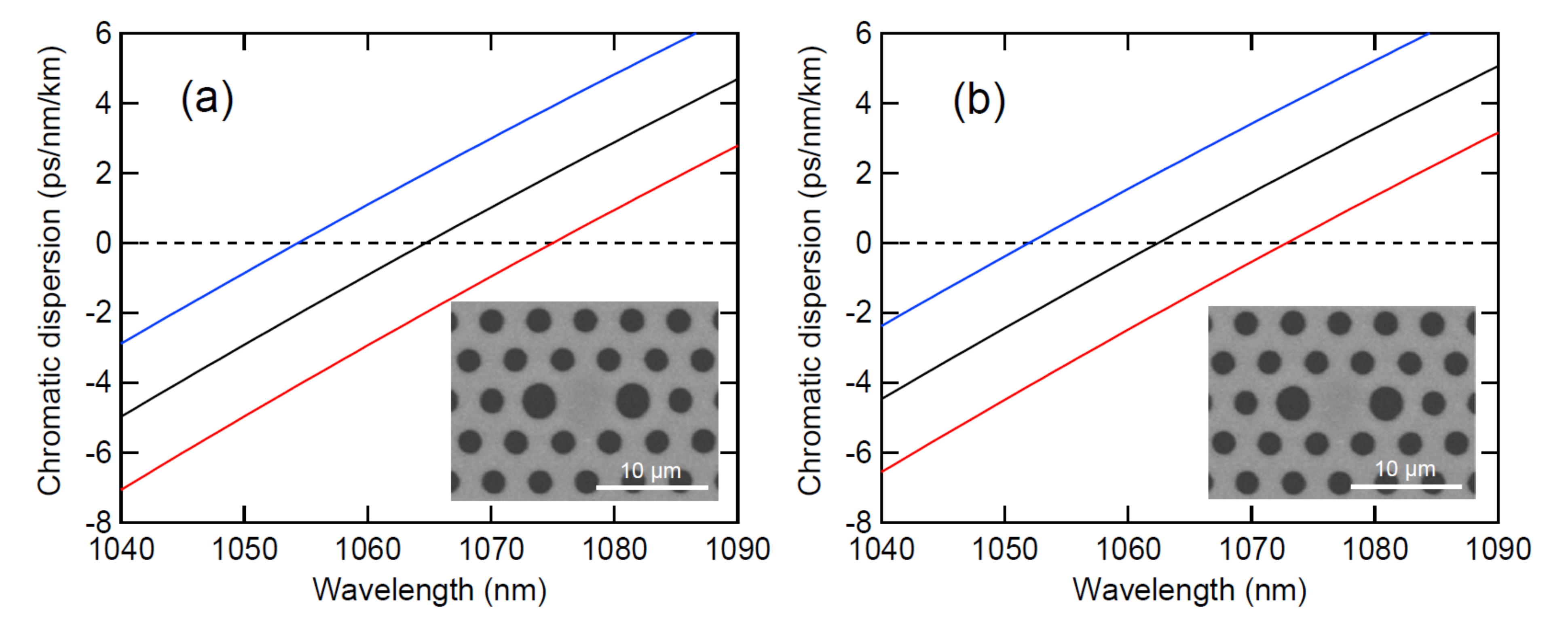}
    \end{center}
     \caption{Simulated dispersion curves of the DOFs used for the soliton (a, DOF\#1) and dispersive shock wave (b, DOF\#2) experiments. Red and blue lines correspond to the maximum and minimum diameters, respectively. Black lines correspond to the average dispersion curves over the whole DOF lengths. Insets show SEM images of the DOF cross sections at the maximum diameters.} \label{figS2}
\end{figure}

Two different DOFs with different lengths and modulation periods have been fabricated for the experiments.
The fiber used for the soliton experiment, labelled DOF\#1 here, is 150 m long and has a modulation period $\Lambda=5$ m. Its outer diameter, which longitudinal evolution is displayed in Fig. 5a of the manuscript, oscillates between 110 and 123 $\mu m$. The inset in Fig. \ref{figS2}a shows a scanning electron microscope (SEM) image of the DOF cross section. It has two bigger holes around the core allowing to increase its form birefringence and to ensure a polarization-maintaining behaviour. Figure \ref{figS2}a shows the full dispersion curve simulated with a commercial finite-element mode solver for the maximum and minimum diameters (red and blue lines respectively), for the neutral axis excited in experiments. The black line represents the average dispersion over the whole DOF length. The average zero-dispersion wavelength is located at 1064.5 nm. Its attenuation was measured to be $0.75$ dB at $1064$ nm and its nonlinear parameter was calculated to be $\gamma=10$ (W km)$^{-1}$ at this wavelength.
The DOF used for the dispersive shock wave experiment, labelled DOF\#2, is 50 m long and has a modulation period $\Lambda=0.5$ m. Its outer diameter (Fig. 5d of the manuscript) oscillates between 110 and 122 $\mu m$. It is based on the same design than DOF\#1 although the geometrical parameters of the cross-section are slightly different. Figure \ref{figS2}b shows the simulated dispersion curves for the maximum and minimum diameters (red and blue lines respectively) as well as the average one (black line). The average zero-dispersion wavelength is located at 1062.5 nm. The overall attenuation is $0.35$ dB at 1064 nm, and its nonlinear parameter is $\gamma=10$ (W km)$^{-1}$ at this wavelength.

\subsection{Numerical simulations and fiber parameters}
The parameters of the fibers that arise from fiber characterization and have been used in numerical simulations are listed in table 1. In the illustrating example reported in Figs. 2 and 3 of the manuscript, only the elements which are essential for the description of the basic phenomenon have been considered. In this case the simulations has been performed by integrating the nonlinear Schr\"odinger equation (NLSE) reported in the method section, i.e. the basic NLSE with additional perturbations due to third-order dispersion $\beta_3$ and the periodic GVD $\delta\beta_2(z)$. For simplicity, the initial conditions have the following hyperbolic secant shapes $E(t,z=0)=\sqrt{P_0}\mathrm{sech}\left(\frac{1.76t}{T_0}\right)$, with $T_0=150$ fs and $P_0=15$ W, for the soliton configuration, and  $T_0=280$ and $P_0=100$ W for the shock wave configuration.
The spectrograms [Fig. 3 in the paper] have been calculated by using a Gaussian pulse as a gate with duration of 1.6 ps.

\begin{table}
	\centering
		\begin{tabular}{l c r}
			Parameter & Soliton & Shock \\ \hline
		    	$\beta_2$ (ps$^2$/km)  & $-1.2$ & $2.9$\\
			$\delta\beta_2$ (ps$^2$/km)  & $1.2$ & $1.2$\\
			$\beta_3$ (ps$^3$/km)  & $0.0716$ & $0.0645$\\
			$\beta_4$ (ps$^4$/km)  & $-1.1\cdot 10^{-4}$ & $-1.1\cdot 10^{-4}$\\
			$\gamma$  (W$^{-1}$km$^{-1}$) & $10$ & $10$\\
			$\alpha$  (dB/km) & $5$ & $7$\\
			$\Lambda$  (m) & $5$ & $0.5$\\
			$L$  (m) & $150$ & $50$
		\end{tabular}
	\caption{Parameters of the fibers employed in the numerical simulations}
	\label{table}
\end{table}

When comparing directly with the experimental results [see Fig. 4c,f in the paper] we make use of the following extended NLSE \cite{agrawal}, which accounts also additional effects such as higher order dispersion effects, Raman effect, self-steepening, and fiber losses
\begin{align} \label{exNLS}
\nonumber \frac{\partial E(z,t)}{\partial z} & = -i\frac{\beta_2(z)}{2}\frac{\partial^2 E(z,t)}{\partial t^2}+\frac{\beta_3}{6}\frac{\partial^3 E(z,t)}{\partial t^3}+i\frac{\beta_4}{24}\frac{\partial^4 E(z,t)}{\partial t^4}-\frac{\alpha}{2}E(z,t)\\&+i\gamma\left(1+i\tau_s\frac{\partial}{\partial t}\right)\times\left(E(z,t)\int{R(t')|E(z,t-t')|dt'}\right),
\end{align}
with $\beta_n$ the $n$th order dispersion terms, $\alpha$ the linear losses, $\tau_s=1/\omega_p$ with $\omega_p$  the central pulsation of the pulse, $R(t)$ the full nonlinear response function that includes the instantaneous (Kerr) and delayed (Raman \cite{hollenbeck}) contributions with fractional weights $f_{Kerr}=0.82$ and $f_{Raman}=0.18$, respectively \cite{agrawal}. Here $\beta_2(z)=\beta_2+\Delta\beta_2\sin(2\pi z/\Lambda)$ and we checked {through numerical simulations} that all other parameters can be assumed to be constant along the fiber length. {Indeed, the modulation of the nonlinear coefficient for instance is about 10~\%, which is about one order of magnitude lower than the $\beta_2$ one \cite{droques}. }

In this case we also use initial conditions that accurately describe the pulses injected in the fiber, which has been experimentally characterized by means of the frequency resolved optical gating (FROG) system. In particular the best fit with FROG data gives slightly chirped input pulses of the following form: $E(t,z=0)=\sqrt{P_0}\mathrm{sech}\left(\frac{1.76t}{T_0}\right)\exp(\frac{-iCt^2}{2T_0^2})$ for the soliton configuration with $T_0=150$ fs and $C=0.35$, and $E(t,z=0)=\sqrt{P_0}\exp \left(\frac{-1.665(1+iC)t^2}{2T_0^2}\right)$ for the shock configuration with $T_0=280$ fs and $C=1.24$. The outcome of the numerical integration of Eq. (\ref{exNLS}) with such initial conditions are directly compared with the experimental results in Fig. 4c,f of the paper.

\begin{figure}[h]
\begin{center}
\includegraphics[width=\textwidth]{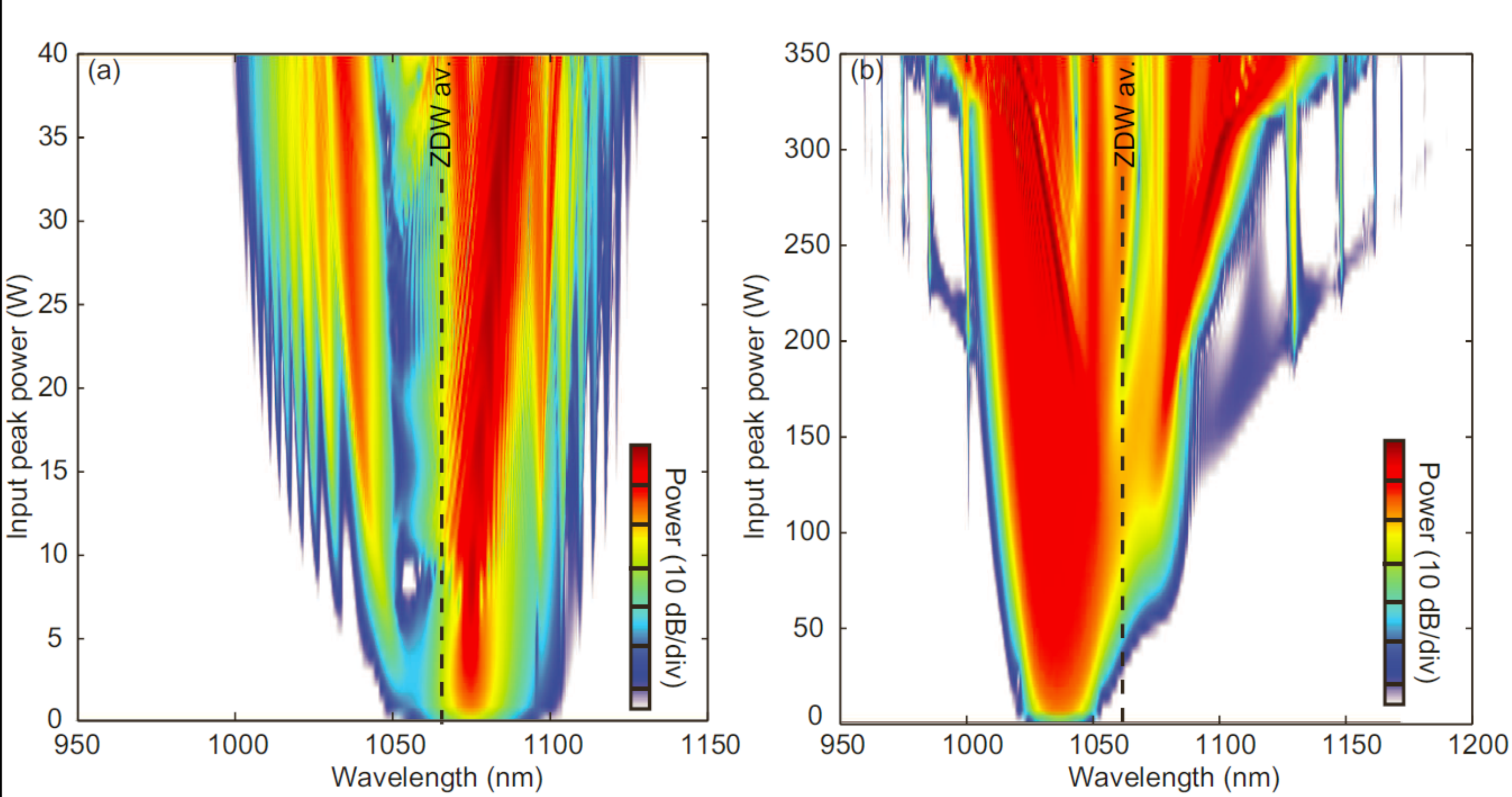}
    \end{center}
     \caption{Colormap of the output spectrum against the input power obtained by means of numerical integration of Eq. (\ref{exNLS}): a. soliton configuration; b. shock configuration. The dashed vertical lines indicates the  average zero dispersion wavelength (ZDW). The maps in panels a and b are directly comparable with the experimental results in Figs. 5b and 5e of the paper, respectively.} \label{figS3}
\end{figure}

Using the realistic simulation parameters given above, we have also simulated the spectral output against pump peak power in both the soliton and dispersive shock wave configuration, respectively. The results, displayed in Fig. \ref{figS3}a,b are directly comparable with the experimental maps reported in Figs. 5b,e of the paper. In both cases the quantitative agreement with experimental data is excellent and confirms the progressive excitation of multiple resonances induced the periodic driving with increasing peak power.

Finally, we emphasize that we have compared the experimental results with the simulations of the full model [Eq. (\ref{exNLS})] in order to have a better accuracy over all the details. However, we have verified that no significant difference arises when integrating the simpler NLSE reported in methods, since the additional terms (steepening, Raman effect, losses, fourth-order dispersion) are indeed negligible in our regime. In particular the parametric excitation of the resonances that we observe in the experiments and in the numerics are indeed quantitatively and accurately explained on the basis of this simpler NLSE.

\clearpage

\end{document}